\documentclass[a4paper,fleqn,usenatbib]{mnras}
\usepackage{newtxtext,newtxmath}
\usepackage[T1]{fontenc}

\DeclareRobustCommand{\VAN}[3]{#2}
\let\VANthebibliography\thebibliography
\def\thebibliography{\DeclareRobustCommand{\VAN}[3]{##3}\VANthebibliography}

\usepackage[dvipsnames]{xcolor}
\usepackage{ae,aecompl}
\usepackage{graphicx}	
\usepackage{amsmath}	
\usepackage{mathtools}
\usepackage{array}
\usepackage{booktabs}
\usepackage{hyperref}
\hypersetup{draft=false}

\title[DEVILS: Satellite Quenching]{Deep Extragalactic VIsible Legacy Survey (DEVILS):  Satellite Quenching at Intermediate Redshift}
\author[L. J. M. Davies]{L. J. M. Davies$^{1}$\thanks{E-mail:
 luke.j.davies@uwa.edu.au},  M. F. Fuentealba-Fuentes$^{1}$, R.  J. Wright$^{1}$, M. Bravo$^{2}$, S. Wagh$^{1,3}$,  M Siudek$^{4,5}$ \\
 $^{1}$ ICRAR, The University of Western Australia, 35 Stirling Highway, Crawley, WA 6009, Australia \\
 $^{2}$ Department of Physics \& Astronomy, McMaster University, 1280 Main Street W, Hamilton, ON, L8S 4M1, Canada \\
 $^{3}$ Australia Telescope National Facility, CSIRO, Space and Astronomy, PO Box 1130, Bentley, WA 6151, Australia \\
 $^{4}$ Instituto de Astrof\'{\i}sica de Canarias, V\'{\i}a L\'actea, 38205 La Laguna, Tenerife, Spain \\
 $^{5}$ Instituto de Astrof\'isica de Canarias (IAC); Departamento de Astrof\'isica, Universidad de La Laguna (ULL), 38200, La Laguna, Tenerife, Spain\\
}

\date{Accepted XXX. Received YYY; in original form ZZZ}

\pubyear{2025}

\begin{document}
\label{firstpage}
\pagerange{\pageref{firstpage}--\pageref{lastpage}}
\maketitle

\begin{abstract}
Determining the processes by which galaxies transition from a star-forming to a quiescent state (quenching) is paramount to our understanding of galaxy evolution. One of the key mechanisms by which this takes place is via a galaxy's interactions with a local, over-dense environment (satellite or environmental quenching). In the very local Universe, we see these processes in action, and can also observe their effects via the distribution of satellite galaxy properties. However, extending similar analyses outside of the local Universe is problematic, largely due to the difficulties in robustly defining environments with small and/or incomplete spectroscopic samples. We use new environmental metrics from the high-completeness Deep Extragalactic VIsible Legacy Survey (DEVILS) to explore the properties of satellite galaxies at intermediate redshift (0.3$<$\textit{z}$<$0.5) and compare directly to the Galaxy And Mass Assembly Survey (GAMA) at 0$<$\textit{z}$<$0.2. Importantly, both the galaxy properties and environmental metrics in DEVILS and GAMA are derived in an identical manner, reducing any methodology biases. We find: i) that satellite galaxies in DEVILS and GAMA show suppressed star-formation in comparison to isolated systems at the same stellar mass, \textcolor{black}{by $\sim$0.5\,dex in log$_{10}$(SFR/M$_{\odot}$\,yr$^{-1}$)}, ii) that this suppression is strongest in higher mass dark matter halos \textcolor{black}{(up to $\sim$1\,dex in log$_{10}$(SFR/M$_{\odot}$\,yr$^{-1}$) in the most massive halos)} and iii) that at fixed stellar and halo mass, this suppression increases with time - with satellite passive fractions increasing by $\sim$10-15\% over the last $\sim$5\,Gyr. This is consistent with previous observations and numerical simulations.                     

\end{abstract}


\section{Introduction}

Galaxies can be classified into two \textcolor{black}{broad types}: blue, gas-rich, star-forming systems, and red, gas-poor quiescent systems with little or no star-formation \citep[$e.g.$][]{Blanton03, Kauffmann03a, Kauffmann04, Baldry04, Balogh04, Brinchmann04, Taylor15, Davies19b}. Over \textcolor{black}{time} the relative proportions of galaxies in each of these \textcolor{black}{types} evolves considerably, with all galaxies initially forming as low stellar mass blue star-forming systems and subsequently evolving until they ultimately reach a quiescent state \citep[$e.g.$][]{Bell04, Faber07, Martin07}. This process is ongoing, with the fraction of galaxies having transitioned from actively star-forming to quiescent increasing with time, until ultimately in the distant future, all galaxies will have reached a passive state, with no active star-formation occurring in the Universe. The process of galaxies transitioning from star-forming to passive is known as `quenching'. Given the ubiquitous nature of quenching and its impact on the galaxy population, studying quenching processes, their frequency, their timescale and their varying impact on subsets of the galaxy population, is paramount for understanding galaxy evolution.  As such, the study of galaxy quenching processes has been one of the key topics in astrophysics for the last 20\,yrs and beyond.                   

Currently both theory and observations suggests two dominant modes of galaxy quenching. First, processes which can occur in all galaxies irrespective of external influence and are correlated with the internal galaxy properties \citep[secular quenching, $e.g.$][]{Kauffmann03b, Driver06,Wake12,Lang14, Barro17, Davies25b}. This mode appears more pronounced at higher stellar masses (log$_{10}$(M$_{*}$/M$_{\odot}$)$>$10.0), with quenched fractions potentially correlating with many properties, such as high central stellar velocity dispersion \citep{Wake12, Teimoorinia16}, the presence of a massive central bulge \citep{Fang13, Bluck14, Bremer18, Cook20} and/or an Active Galactic Nucleus \citep[AGN, $e.g.$][]{Nandra07, Davies25b}. However, in order to reproduce the observed distribution of low-mass systems, simulations also must contain gas outflows which inhibit star-formation in the lowest mass galaxies \citep[$e.g.$ log$_{10}$(M$_{*}$/M$_{\odot}$)$<$9.0,][]{Dekel86}. These outflows are generally attributed to stellar feedback processes \citep{DallaVecchia08}, which at log$_{10}$(M$_{*}$/M$_{\odot}$)$\gtrsim$9.0 do not drive the gas with enough energy to escape the galaxy's gravitational potential and therefore, cannot lead to quenching  \citep[$e.g.$][]{Dekel86}. This indicates that secular quenching is likely bimodal with stellar mass - affecting both low and high-mass galaxies but leaving intermediate mass systems relatively unscathed \citep[$e.g.$ see][]{Katsianis19, Davies19a, Davies22, Davies25b}.

The second quenching mode is driven by a galaxy's local environment. Over-dense environments such as clusters, groups or even close pairs \citep[see][]{Patton11, Robotham14, Davies16a} can either remove or inhibit the supply of gas required for ongoing star formation, leading to quenching \citep[$e.g.$][]{Peng10, Peng12, McGee14, Haines15, Schaefer17, Bluck21a, Bluck21b, Cortese21, Cleland21}. Many physical processes can drive this, such as tidal and ram-pressure stripping \citep{Gunn72, Moore99, Poggianti17,Brown17, Barsanti18}, starvation/strangulation \citep{Larson1980, Moore99, Peng15, Nichols11},  and/or harassment \citep{Moore96}. Environmental quenching is more likely to occur in intermediate-to-low-mass galaxies \citep[log$_{10}$(M$_{*}$/M$_{\odot}$)$<$11.0, $e.g.$ see][]{Davies19a}, and is found to correlate with local galaxy density within groups/clusters \citep{Peng12, Treyer18}, and group/cluster-centric position \citep{Wolf09, Wetzel12, Woo15, Barsanti18, Oxland24} - likely due to the fact that low-mass galaxies moving through over-dense environments cannot retain,  \textcolor{black}{and/or cool the gas required for star-formation}. 

\textcolor{black}{Importantly, this environmental quenching should only affect satellites ($i.e.$ galaxies within a common dark matter halo, which do not reside at the centre of the halo). The most massive galaxies that sit at the centre of their haloes are likely not subject to stripping, etc, and have minimal tidal interactions/harassment, such that they can retain their gas}. Thus, we may expect centrals and satellites to undergo different quenching mechanisms and display different passive fractions when controlled for all other effects \citep[$e.g.$][]{vandenBosch08, Weinmann09, Wetzel12, Peng12, Knobel13,Robotham14, Grootes17}. This is the typically accepted model for environmental quenching processes, where satellite galaxies undergo additional quenching in over-dense environments \citep[$e.g.$][]{Wetzel13, Treyer18, Davies19b}, especially when a satellite is significantly less massive than it's central/halo. This model is both used in numerous galaxy evolution models \citep[$e.g.$][]{Cole00, Henriques15, Cora18, Lagos18} and is observed in hydrodynamic simulations \citep[$e.g.$][]{Bahe15, Zinger18, Lotz19, Wright19, Wright22, Donnari21a, Donnari21b}.

Observationally, the majority of studies probing environmental quenching mechanisms in large statistical samples are undertaken at a single epoch at low redshift \citep[$e.g.$][]{Peng12, Davies19b, Bluck21b, Oxland24}. This is largely due to the observational complications with robustly defining galaxy environments outside of the local Universe. Ideally, this requires high completeness spectroscopic samples with robust and precise redshifts with which to link galaxies to common dark matter halos and to measure dark matter halo masses from galaxy velocity dispersions \citep[$e.g.$][]{Yang07, Robotham11, Tempel12}. Until recently there has been a paucity of similar high completeness samples in the more distant Universe, rendering the measurement of environment in large galaxy samples, particularly in terms of dark matter halo mass, problematic. However, there have been a number of studies aiming to overcome these issues to explore satellite quenching at earlier times. For example, works such as \cite{Peng10} and \cite{Fossati17} \textcolor{black}{match galaxies} to mock galaxy simulations to determine probabilistic halo masses. Using this approach \cite{Fossati17} find that, at a fixed halo mass, satellite passive fractions are lower in the distant Universe, and that the fraction of passive satellites is increasing with time, suggesting we are witnessing environmental quenching mechanisms in action. Other studies perform a complementary approach of studying satellite quenching in the most over-dense cluster environments at higher redshifts \citep[$e.g.$][]{Muzzin14, Haines15, Foltz18, vanderBurg20, Turner21, Mao22,Siudek22, Figueira24}  finding that these environments have strong impact on the star-forming properties of their satellite galaxies, suggesting that satellite galaxy quenching has been a ubiquitous feature of galaxy evolution for at least the last 10\,Gyrs. 

However, comparing these studies across large evolutionary baselines in a robust and consistent manner is challenging. These studies use vastly different techniques to determine their environmental metrics, often with large assumptions and/or conversions from indirect observables, or target specific individual halos which may be atypical of the general population. They also use different methodologies for measuring star-formation rates (SFRs) and stellar masses, and different techniques for selecting passive/star-forming galaxies. As such, they are not easily comparable when attempting to explore the time evolution of satellite quenching processes.     

Here we aim to overcome some of these issues by using a combination of the Deep Extragalactic VIsible Legacy Survey \citep{Davies18, Davies21} and Galaxy And Mass Assembly Survey \citep[GAMA][]{Driver11, Driver16, Liske15, Baldry18}. These surveys both provide deep, high spectroscopic completeness sample with which to study galaxy and environmental evolution. GAMA provides a large area sample in the local Universe ($z$$<$0.2), while DEVILS provides a comparable, small area, sample extending out to $z$$<$0.8. Importantly, DEVILS was designed to match GAMA in a number of respects, but to extend out to the more distant Universe - allowing us to define comparable samples of galaxies across a broad evolutionary range.  Within these samples we now also have galaxy properties (stellar masses, SFRs, etc) and environmental diagnostics (halo mass, etc) derived in an identical manner using the same codes and techniques. This allows us to probe galaxy evolution process in both surveys simultaneously, while minimising methodology biases and inconsistencies. In this work we will use the DEVILS and GAMA sample in combination to explore the evolution of satellite quenching over the last $\sim$5\,Gyr.  Throughout this paper we use a standard $\Lambda$CDM cosmology with {H}$_{0}$\,=\,70\,kms$^{-1}$\,Mpc$^{-1}$, $\Omega_{\Lambda}$\,=\,0.7 and $\Omega_{M}$\,=\,0.3.

\section{Data and Sample Selection}

In this work we use samples derived from both the Deep Extragalactic VIsible Legacy Survey (DEVILS) at intermediate redshift and the Galaxy And Mass Assembly Survey (GAMA) at low redshift. While the data products used in this work are described extensively elsewhere, we briefly summarise the key measurements here.  

\subsection{The Deep Extragalactic VIsible Legacy Survey}

DEVILS is a spectroscopic survey undertaken at the Anglo-Australian Telescope (AAT), which aimed to build a high completeness ($>$85\%) sample of $\sim$50,000 galaxies to Y$<$21\,mag in three well-studied deep extragalactic fields: D10 (COSMOS), D02 (ECDFS) and D03 (XMM-LSS). The survey aims to provide the first high completeness sample at 0.3$<$$z$$<$1.0, allowing for the robust parametrisation of group and pair environments in the distant Universe.  DEVILS also serves as a precursor to the Wide Area VISTA Extragalactic Survey \citep[WAVES,][]{Driver19} deep program, which will cover similar galaxy populations, and use similar methodologies to derive galaxy and environmental properties, but will cover $\sim$15 times the volume.  The science goals of both DEVILS and WAVES deep are varied, from the environmental impact on galaxy evolution at intermediate redshift, to the evolution of the halo mass function over the last $\sim$7\,billion years. For full details of the DEVILS survey science goals, survey design, target selection, photometry and spectroscopic observations see \cite{Davies18, Davies21}. Full details of the DEVILS spectroscopic sample will be presented in Davies et al (in prep).
   
The DEVILS regions were chosen to cover areas with extensive exisiting and ongoing imaging to facilitate a broad range of science. In this work we only use the DEVILS D10 field which covers the Cosmic Evolution Survey region \citep[COSMOS,][]{Scoville07}, extending over 1.5\,deg$^{2}$ of the UltraVISTA \citep{McCracken12} field and centred at R.A.=150.04, Dec=2.22. This field is prioritised for early science as it is the most spectroscopically complete, has the most extensive multi-wavelength coverage of the DEVILS fields,  has already been processed to derive robust galaxy properties through spectral energy distribution (SED) fitting, and has had robust environmental metrics defined (see below). 

\subsubsection{Stellar Masses and SFRs}

For DEVILS galaxy properties, in this work we use the outputs of the SED fitting process outlined in \cite{Thorne21} and \cite{Thorne22}. Briefly, \cite{Thorne21} fits galaxies in the D10 region using the \textsc{ProSpect} \citep{Robotham20} SED fitting code to estimate galaxy properties such as stellar mass, SFR, star-formation history (SFH) and metallicity. In \cite{Thorne22}, this process is updated to include an AGN model, which allows for the identification of sources hosting bright AGN and improvements to the other derived properties for AGN host galaxies. While the overwhelming majority of sources in the D10 sample do not change their properties, the sources identified as AGN do, in some case, have significant changes, particularly to their SFR and SFH (as UV and MIR-FIR is now attributed to the AGN and not star-formation). A detailed description of how galaxy properties in the D10 sample are affected by the inclusion of the AGN is included in \cite{Thorne22}, so we refer the reader to that work. However, for completeness we briefly note that these works use photometric data described in \cite{Davies21} covering 22 bands from the FUV to FIR. We note briefly here, that for the DEVILS sample used in this work, 99\% are detected $>$2$\sigma$ in UV bands, 100\% in the optical, 100\% in the NIR, 100\% in the MIR and 60\% in the FIR. As such, the SED-derived properties used here are likely robust. In this work we use the best-fit stellar mass and SFRs taken from the \textsc{ProSpect} fits, which contain a possible AGN component (which likely have better defined properties). 

\subsubsection{Groups and Satellites}

For DEVILS environmental properties, we use the D10 group catalogues derived in Bravo et al (in prep). This work uses a the same approach as outlined in \cite{Robotham11} based on a bespoke friends-of-friends based grouping algorithm, which was tested extensively on mock GAMA galaxy light cones.  Bravo et al (in prep) optimises friends-of-friends linking lengths for DEVILS specifically and tests the success and failures of group recovery using updated DEVILS mock galaxy light cones from the \textsc{shark} semi-analytic model \citep{Lagos18, Lagos24}. First, galaxies are linked to common halos and assigned a group ID and then group halo masses are measured from the galaxy kinematics, corrected to remove biases when compared to the mock galaxy light cones. Here we use the scaled mass proxy, M$_{\mathrm{halo}}$$\sim$AR$_{50}\sigma^2$. Where $R_{50}$ is the radius containing 50\% of the group members, $\sigma$ is the group velocity dispersion and $A$ is a functional scaling factor based on group multiplicity and redshift \citep[see Section 4.3 of ][]{Robotham11}. We note here at the DEVILS group catalogues extend to multiplicity N=2 systems (pairs). However, in all of our subsequent analysis we limit the sample to N$>$2 systems - where halo masses are better constrained as there are more tracers of halo velocity dispersions.  

The DEVILS group catalogue gives a number of different central/satellite assignments based on various approaches, such as the most luminous group member (brightest group galaxy, BGG). Here we use the group galaxy with the largest stellar mass as the group central. All other group galaxies are classed as `satellites'. Spectroscopically confirmed DEVILS galaxies which are not in a N>1 group are classed as `isolated centrals'.  For full details of the group finding and halo mass estimates, see \cite{Robotham11} and Bravo et al (in prep).  

\begin{figure*}
\begin{center}
\includegraphics[scale=0.7]{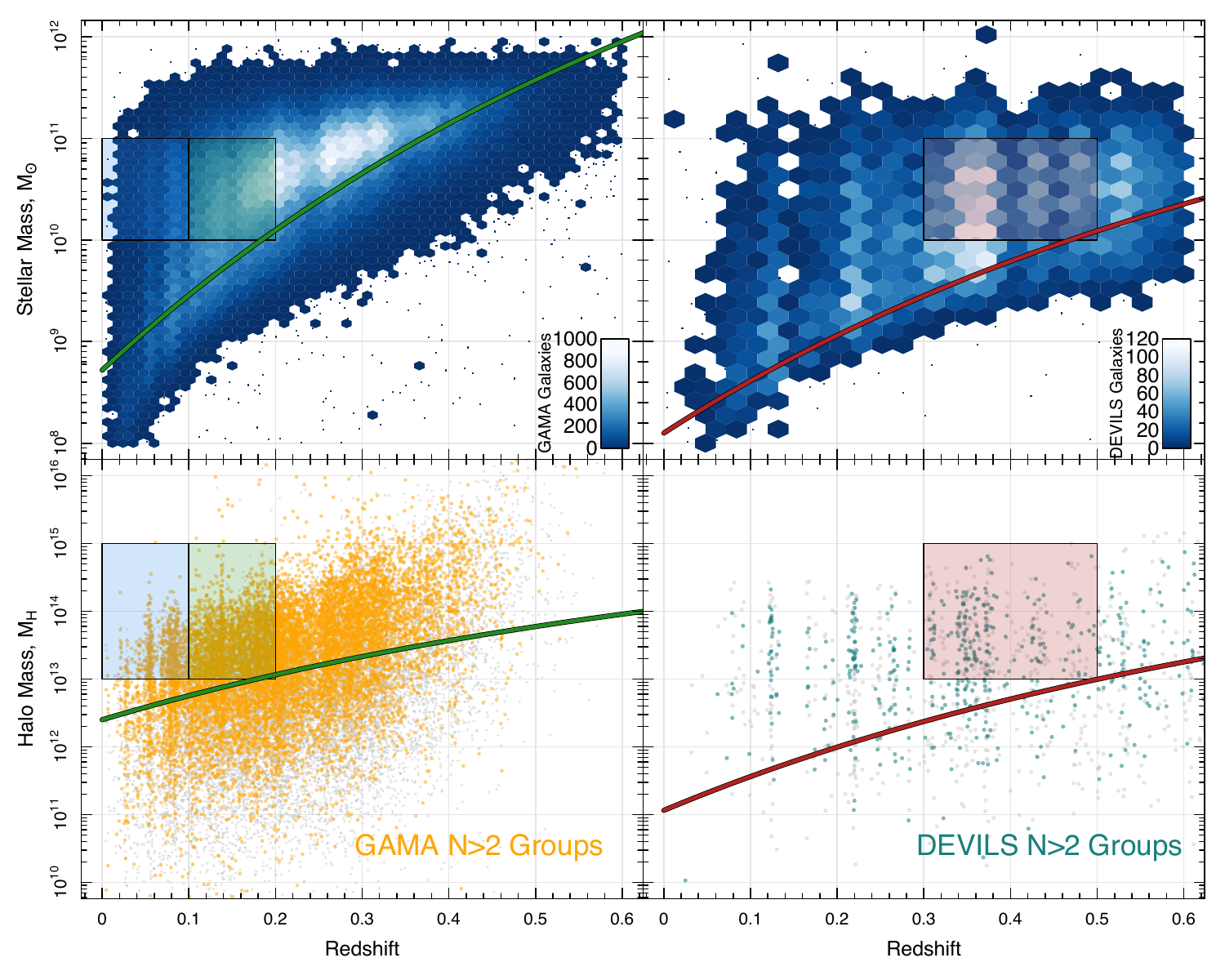}

\caption{The redshift-mass distribution of galaxies (top row) and dark matter halos (bottom row) in GAMA (left column) and DEVILS (right column). All properties (Stellar mass, redshift, halo mass) are derived using the same method in both GAMA and DEVILS. Stellar mass, halo mass and redshift ranges used in this work are shown in coloured rectangles. Green and red lines show the stellar and halo mass completeness limits for GAMA and DEVILS respectively.  DEVILS at 0.3$<$$z$$<$0.5 provides a similar sample in terms of stellar mass and halo mass to GAMA at $z$$<$0.2 allowing a direct comparison  of the time evolution of properties since $z$$\sim$0.5. In our subsequent analysis we split the GAMA sample into two redshift bins, indicated by the different coloured rectangles. }
\label{fig:sample}
\end{center}
\end{figure*}

\subsection{The Galaxy And Mass Assembly Survey}

The GAMA survey fourth data release covers 286\,deg$^{2}$ to a main survey limit of $r_{\mathrm{AB}}$$<$19.8\,mag in three equatorial (G09, G12 and G15) and two southern (G02 and G23 - survey limit of $i_{\mathrm{AB}}$$<$19.2\,mag in G23) regions. The spectroscopic survey was undertaken using the AAOmega fibre-fed spectrograph \citep[][]{Saunders04,Sharp06} in conjunction with the Two-degree Field \citep[2dF,][]{Lewis02} positioner on the Anglo-Australian Telescope, and obtained redshifts for $\sim$240,000 targets covering 0$<$$z$$\lesssim$0.5 with a median redshift of $z$$\sim$0.2, and highly uniform spatial completeness \citep[see][for a summary of GAMA observations]{Baldry10,Robotham10,Driver11}. Full details of the GAMA survey can be found in \citet{Driver11, Driver16},  \citet{Liske15} and \citet{Baldry18}. In this work we limit our sample to galaxies that have a confirmed, local-flow-corrected redshift at 0.01$<$$z$$<$0.2. 

\subsubsection{Stellar Masses and SFRs}

For GAMA galaxy properties, we also use the outputs of the \textsc{ProSpect} SED fitting process undertaken in an identical manner to DEVILS. This process is initially outlined in \cite{Bellstedt20b} but was subsequently expanded to use to include an AGN model, the same SFH choices and methodology choices outlined in \cite{Thorne22}. The GAMA  \textsc{ProSpect} fitting is undertaken using photometry covering similar rest-frame emission wavelengths as DEVILS, with photometric data outlined in \cite{Bellstedt20a, Bellstedt20b}. From this work we use best-fit stellar mass, SFRs and SFHs in an identical manner to the DEVILS sample described above. As such, this likely removes any significant biases induced by choice of methodology in the derived galaxy properties. The only small caveat to this is the fact that GAMA galaxies exist at the different redshift to DEVILS (and as such has slightly different rest-frame coverage) and have different depth input photometry (and as such may have larger errors in the fitting). However, these differences will likely result very minor differences in the efficacy of the SED fitting. Once again, for completeness, we note briefly here that for the GAMA sample used in this work, 99\% are detected at $>$2$\sigma$ in UV bands, 100\% in the optical, 100\% in the NIR, 100\% in the MIR and 68\% in the FIR.

\subsubsection{Groups and Satellites}

For our GAMA environmental properties we use the G$^{3}$C catalogue which includes the identification of all galaxy groups and pairs within GAMA \citep[which is described in][but now cover all GAMA regions. See also \citealp{Robotham12,Robotham13,Robotham14, Davies15b, Davies19b}]{Robotham11}. As in DEVILS, the GAMA group catalogue is also produced using a bespoke friends-of-friends based grouping algorithm, which was tested extensively on mock GAMA galaxy light cones, and assigns $\sim$40\% of GAMA galaxies to multiplicity N$>$1 pairs and groups. As with DEVILS, we define a group as a system with multiplicity N$>$2. Once again, for halo masses we also use scaled mass proxy, M$_{\mathrm{halo}}$$\sim$AR$_{50}\sigma^2$ calculated in the same manner as  DEVILS. Finally, we also use the group galaxy with the largest stellar mass as the group central. All other group galaxies are classed as `satellites'. Spectroscopically confirmed GAMA galaxies which are not in a N>1 group are classed as `isolated centrals'. 

\vspace{2mm}

\textcolor{black}{Before proceeding, we also briefly note the suitability of both GAMA and DEVILS in exploring the environmental impact on satellite galaxies. The key ingredients required to undertake this science are robust environmental metrics and central/satellite assignments, highly complete samples of satellites at a fixed stellar mass, and robust stellar masses and SFRs. For environmental metrics, GAMA and DEVILS are both specifically designed to produce highly spatially- and stellar mass-complete spectroscopic samples extending to low stellar masses. For example, GAMA extends to $\sim$2\,dex lower in stellar mass than the Sloan Digital Sky Survey \citep[SDSS, $e.g.$][]{Abazajian09} and is more complete to closely separated galaxies, $>$95 per cent compared to $>$70 per cent \citep[see][]{Liske15}.  Likewise, DEVILS extends to $\sim$1\,dex lower in stellar mass than similarly complete samples in COSMOS, and has much higher spectroscopic completeness on small spatial scales (Davies et al , in prep). Essentially, both surveys will identify more galaxies per dark matter halo than other comparable surveys, allowing us to better constraint dark matter halo masses \citep[see figure 6 of][for a direct comparison between GAMA and SDSS group multiplicities for the same haloes]{Davies19b}. In terms of galaxy properties, both GAMA and DEVILS undertake the same SED fitting procedure using over $>$20\,band spanning the UV-FIR for sources with robust spectroscopic redshifts. This leads to highly-robust measurements of stellar mass and SFR, and well-constrained errors on both properties \citep[see][]{Bellstedt20b,Thorne22}. In combination this allows for the identification and parametrisation of both lower-mass groups and their satellite populations. Coupled with the ability to define minimally-biased samples spanning the last $\sim$5\,Gyrs across DEVILS and GAMA, this makes these surveys ideally suited to explore the evolution of satellite quenching. }

\subsection{Sample selection}
\label{sec:select}

Using the DEVILS and GAMA samples defined above, we now wish to select subsamples in both stellar mass-redshift and halo mass-redshift which can be directly comparable across multiple epochs. Figure \ref{fig:sample} displays the full sample of GAMA (left column) and DEVILS (right column) spectroscopically confirmed galaxies used in this analysis as a function of redshift. The top row shows the stellar mass distribution, while the bottom row shows the halo mass distribution.   

First, to define a stellar mass complete sample, we use a similar approach outlined in previous works for GAMA \citep[$e.g.$][]{Davies16a} and DEVILS \citep[$e.g.$][]{Thorne21, Fuentealba25}. In bins of lookback time, we determine the distribution rest-frame g-i colour of galaxies and find the stellar mass limit which encompasses $90\%$ of the colour distribution at a given lookback time. We then linearly fit the stellar mass limit vs lookback time relation to obtain a smoothly evolving stellar mass limit with redshift. This is undertaken on GAMA and DEVILS separately and is displayed in the top row of Figure \ref{fig:sample} as the green and red lines respectively. Above these lines our samples are largely complete in stellar mass to both red and blue galaxies. We note that these completeness limits are consistent with previous GAMA and DEVILS papers. We then define a stellar mass range where both GAMA and DEVILS are complete, spanning 0$<$$z$$<$0.5 ranging from 10$<$log$_{10}$(M$_{\star}$/M$_{\odot}$)$<$11. Below this the sample would be incomplete to red galaxies, and above this the volumes are too low to accurately probe the galaxy population. In GAMA we split into two redshift ranges, as we have adequate sample sizes to subdivide the population. In DEVILS we retain a single redshift range as sample sizes are smaller (this also a results in bins which are similar width in lookback time). These selection regions are highlighted as coloured polygons in the top row of figure \ref{fig:sample}.

Following this, we then define a halo-mass-complete sample over the same redshift ranges. Defining the halo mass at which a sample is complete at a given redshift is more problematic. Here we simply take the full sample of N$>$1 groups in bins of lookback time. We then identify where the halo number counts turn-over at given lookback time, and set this point as our halo mass incompleteness limit. We then linearly fit the halo mass limit vs lookback time relation to obtain a smoothly evolving halo mass limit in redshift. This is again undertaken on GAMA and DEVILS separately and is displayed in the bottom row of Figure \ref{fig:sample} as the green and red lines respectively. We then identify that our halo mass samples at each epoch are complete  at  log$_{10}$(M$_{\mathrm{halo}}$/M$_{\odot}$)$>$13 (shown as the coloured polygons in the bottom row of Figure \ref{fig:sample}). 

Samples of galaxies and environments contained within these coloured polygons will be used in subsequent analyses to explore the time evolution of satellite quenching.                  

\begin{figure*}
\begin{center}
\includegraphics[scale=0.16]{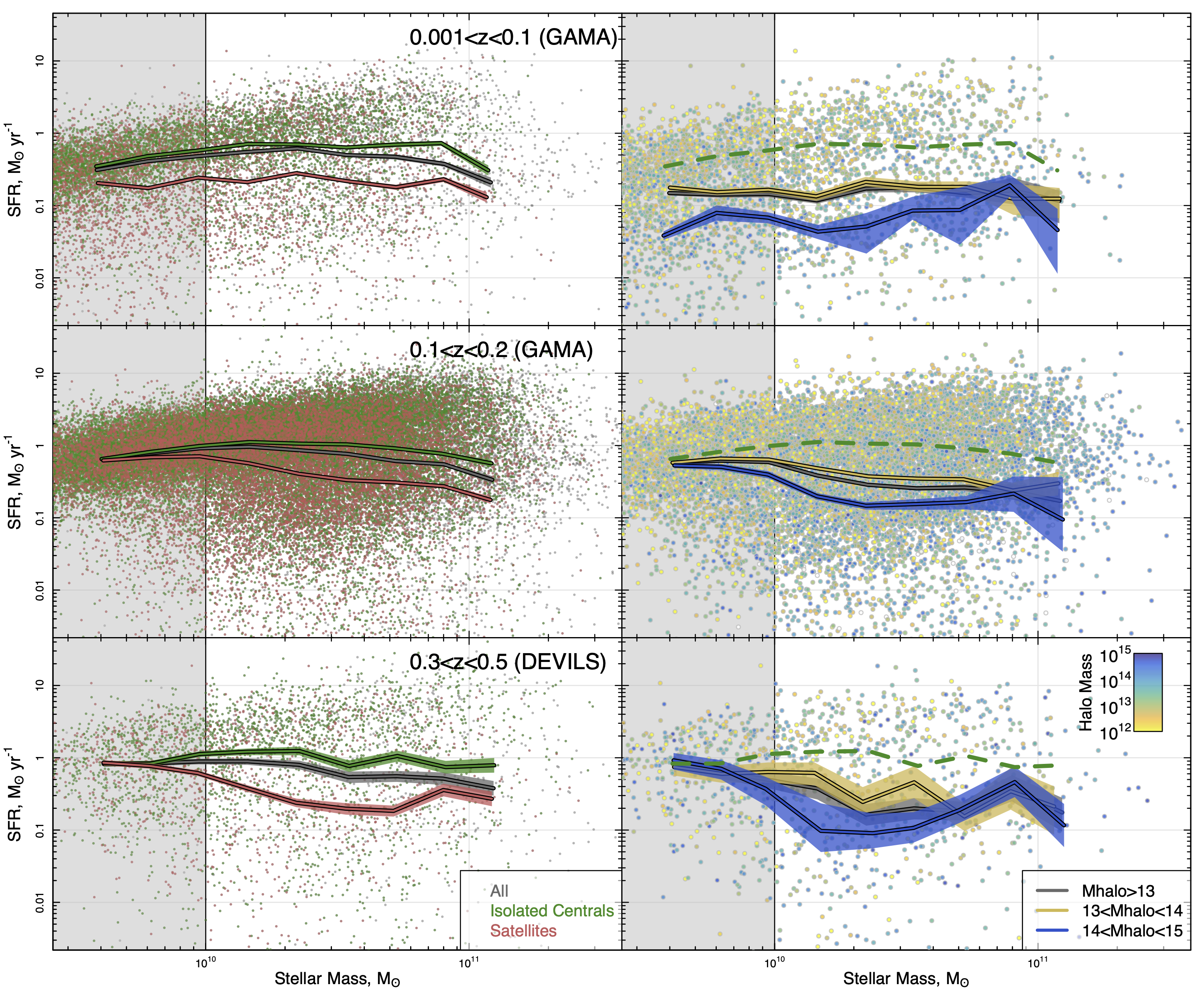}
\vspace{-3mm}

\caption{The SFR-M$_{\star}$ plane at the three different redshifts probed in this work. Top two rows are taken from GAMA, bottom row from DEVILS. The left panel shows all galaxies in the sample. Grey points show all galaxies, green points are isolated central galaxies, while red points show satellite galaxies. The lines display the running median of SFR for each sample. The right panel shows just the satellite galaxies coloured by halo mass.  Lines display the running median of SFRs for all log$_{10}$(M$_{\mathrm{halo}}$/M$_{\odot}$)$>$13 halos (grey), high mass 14$<$log$_{10}$(M$_{\mathrm{halo}}$/M$_{\odot}$)$<$15 halos (blue) and low mass 13$<$log$_{10}$(M$_{\mathrm{halo}}$/M$_{\odot}$)$<$14 halos (gold).  \textcolor{black}{Error polygons are defined using 5,000 Monte-Carlo realisations of the data when randomly sampling data points within their SFR and stellar mass errors (see text for details).}  The grey shaded region in all panels shows where our sample becomes incomplete in stellar mass.}
\vspace{-3mm}

\label{fig:SFS}
\end{center}
\end{figure*}

\begin{figure*}
\begin{center}
\includegraphics[scale=0.58]{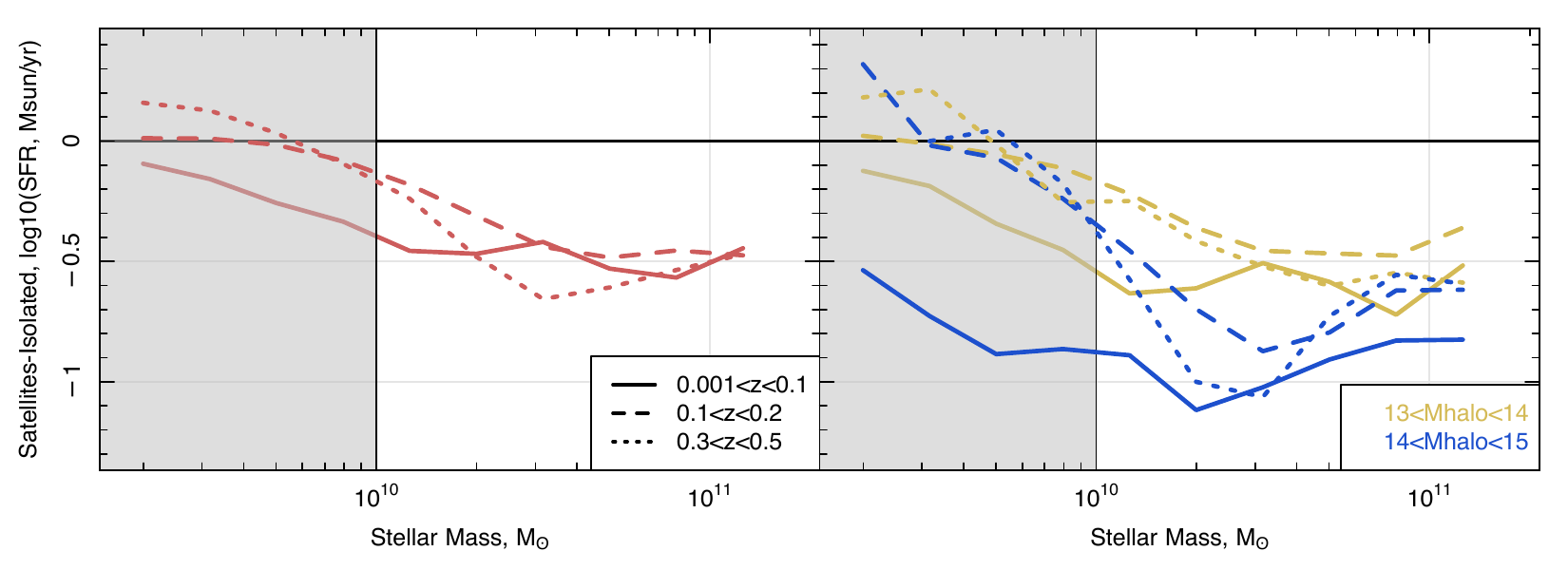}
\vspace{-3mm}

\caption{The log$_{10}$(SFR/M$_{\odot}$\,yr$^{-1}$) offset between the running median SFR for isolated central and satellite galaxies. The left panel shows the offset for all satellites at the three redshifts probed in this work, while the right panel shows the same but sub-divided into two halo mass ranges. The grey shaded region is where our samples become incomplete in stellar mass. }     
\vspace{-3mm}
 
\label{fig:SFSoffset}
\end{center}
\end{figure*}

\section{Analysis and Discussion}

\subsection{SFR Suppression in Satellite Galaxies}
\label{sec:suppression}

First, we explore the SFRs of all satellite galaxies \textcolor{black}{(irrespective of halo mass)} in comparison to isolated centrals as a function of stellar mass. To do this, the left column of Figure \ref{fig:SFS} shows the SFR-M$_{\star}$ plane at each of the redshifts probed in this work (rows). We colour points identified as a satellite (in all N$>$2 halos) in red and isolated centrals in green. Note, the remaining non-green or -red points are the centrals of N$>$2 halos or galaxies in a N=2 system (which we do not consider here). The grey shaded region shows our stellar mass completeness limit and as such, below this samples are incomplete and therefore unreliable. 

We then plot the running median of SFRs as a function of stellar mass for all galaxies (grey), isolated centrals (green) and satellite galaxies (red). \textcolor{black}{To determine likely error ranges for these running medians, we use the upper and lower 1$\sigma$ bounds of the \textsc{ProSpect}-fitting MCMC chains in both SFR and stellar mass. We then run 5,000 Monte-Carlo realisations of the sample, where in each realisation we sample the SFR and stellar mass within the upper and lower bounds, and recalculate the running median. We then calculate the 5\%-95\% quantile range of all realisations, as the possible spread in running median, incorporating both errors in SFR and stellar mass. These ranges are shown as coloured polygons in Figure \ref{fig:SFS}. We note here, that in this Monte-Carlo process we also calculate passive fractions for each realisation, which will be used later in this work}.
 
At stellar masses, above our completeness limit, we find that satellite galaxies have systematically lower SFRs than isolated centrals \textcolor{black}{even when considering the error analysis described above}, and that this offset is largely the same at all epochs. This suggests that over-dense environments have been impacting satellite galaxies over the last $\sim$5\,Gyr (since $z$$\sim$0.5) and that this impact is likely similar in how it affects star-formation over this range of time.

However, care must be taken here, as this analysis includes all satellite galaxies, irrespective of halo mass. To expand upon this, in the right column of Figure \ref{fig:SFS}, we now only show satellite galaxies, with points coloured by log$_{10}$(M$_{\mathrm{halo}}$/M$_{\odot}$). We find that there is a qualitative trend of galaxies in higher mass halos have lower SFRs at a given stellar mass (points become more blue as we move to lower SFRs). To highlight this further, we take three halo mass ranges above our halo mass completeness limit defined in Section \ref{sec:select} and plot the running  median of SFRs as a function of stellar mass for each range \textcolor{black}{with error polygons calculated using the same approach as above}. First, we show satellites in all log$_{10}$(M$_{\mathrm{halo}}$/M$_{\odot}$)$>$13 halo in grey, and then subdivide this into 13$<$log$_{10}$(M$_{\mathrm{halo}}$/M$_{\odot}$)$<$14 halos (gold) and 14$<$log$_{10}$(M$_{\mathrm{halo}}$/M$_{\odot}$)$<$15 halos (blue). We also over-plot the isolated centrals line from the left panels as the dashed green line for comparison. 

We find that satellites in all halos above our halo completeness limit are suppressed in star-formation at all stellar masses above our stellar mass completeness limit. What is more, we find that this effect is more significant for higher mass halos at all redshifts\textcolor{black}{, even when taking into account possible errors. This is only weakly true for the DEVILS sample, when considering the error polygons. We also find that this effect is strongest at 10$<$log$_{10}$(M$_{\mathrm{\star}}$/M$_{\odot}$)$<$10.5, where above this the differences between halos mass ranges are negligible. However,}  this strongly suggests it is the effect of environment that is leading to the suppression in star-formation.

\begin{table*}
\caption{Median satellite SFR offset from the isolated galaxy population at 10$<$log$_{10}$(M$_{\mathrm{\star}}$/M$_{\odot}$)$<$11 in log$_{10}$(SFR/M$_{\odot}$\,yr$^{-1}$). Error ranges are calculated using the range of offsets in the Monte-Carlo realisations described in the text. }
\begin{center}
\begin{tabular}{cccccc}
& & \multicolumn{4}{c}{Median satellite SFR offset from isolated galaxies, log$_{10}$(SFR/M$_{\odot}$\,yr$^{-1}$)} \\
\hline
Sample & $z$ & All & log$_{10}$(M$_{\mathrm{halo}}$/M$_{\odot}$)$>$13 & 13$<$log$_{10}$(M$_{\mathrm{halo}}$/M$_{\odot}$)$<$14 & 14$<$log$_{10}$(M$_{\mathrm{halo}}$/M$_{\odot}$)$<$15 \\ 
\hline
GAMA & 0.01$<$$z$$<$0.1 & 0.63 $\pm$ 0.05 & 0.84 $\pm$ 0.07 & 0.76 $\pm$ 0.06 & 1.12 $\pm$ 0.08  \\
GAMA & 0.1$<$$z$$<$0.2 & 0.52 $\pm$ 0.01 & 0.69 $\pm$ 0.02 & 0.55 $\pm$ 0.02  & 0.98 $\pm$ 0.05   \\
DEVILS & 0.3$<$$z$$<$0.5 & 0.58 $\pm$ 0.10 & 0.58 $\pm$ 0.14 & 0.46 $\pm$ 0.18  & 0.85 $\pm$ 0.17  \\
\end{tabular}
\end{center}
\label{tab:values}
\end{table*}

\begin{figure}
\begin{center}
\includegraphics[scale=0.55]{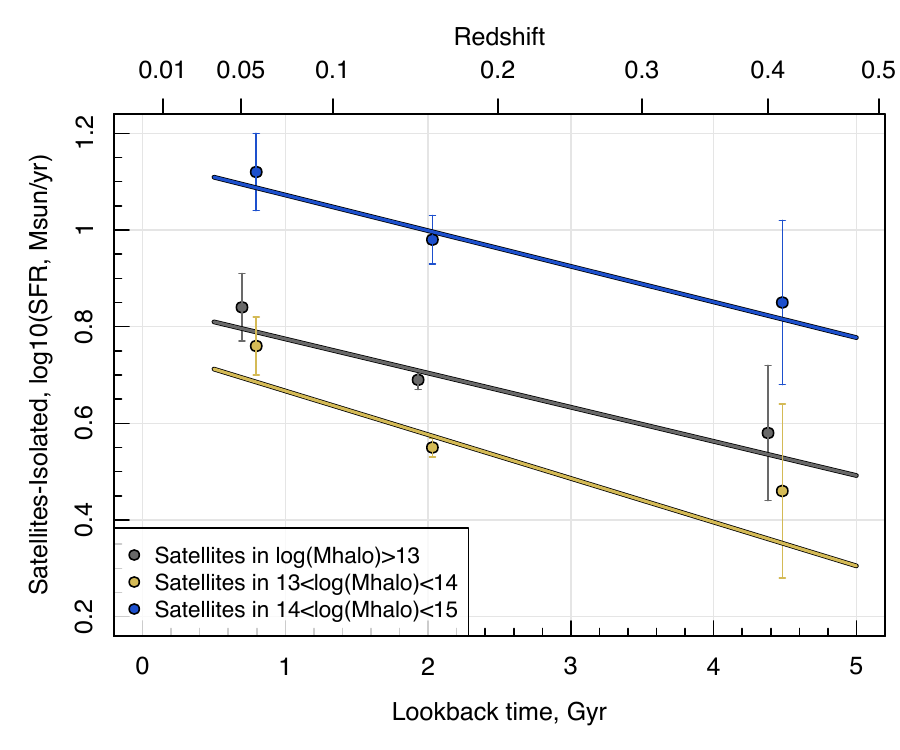}

\caption{The  evolution of the median log$_{10}$(SFR/M$_{\odot}$\,yr$^{-1}$) offset between isolated and satellite systems at 10$<$log$_{10}$(M$_{\mathrm{\star}}$/M$_{\odot}$)$<$11, split in the the halo mass ranges used in this work. Linear regression fits are shown as solid lines to highlight the observational trends.}
\label{fig:timeOffset}
\end{center}
\end{figure}

We also note that the difference between isolated centrals and satellites appear similar at all epochs, potentially suggesting that the impact of environment on star-formation in satellites is largely similar at all epochs probed here (when controlled for halo mass and stellar mass).  To further highlight this, in Figure \ref{fig:SFSoffset} we show the offset, in log$_{10}$(SFR/M$_{\odot}$\,yr$^{-1}$), between the running median for satellites and isolated centrals at each epoch. The left panel displays all satellites, while the right panel shows only those in the two halo mass ranges defined above.\textcolor{black}{We opt not to show the error polygons in this figure as it is then difficult to see the overall trends. Here, we simply wish to show the similarity between trends at each epoch when using best-fit properties}. For all satellites we find that there is a consistent $\sim$0.5-0.6\,dex log$_{10}$(SFR/M$_{\odot}$\,yr$^{-1}$) offset between isolated central and satellite galaxies at all redshifts probed here. \textcolor{black}{Details of the systematic SFR offsets between isolated centrals and satellites, calculated over the full 10$<$log$_{10}$(M$_{\mathrm{\star}}$/M$_{\odot}$)$<$11 range, are given in Table \ref{tab:values}. These offsets} suggest that when controlled for stellar mass, satellite galaxies show a largely consistent suppression of star-formation across the last $\sim$5\,Gyr. We do see tentative hints that the lowest redshift bin shows more suppression of star-formation than higher redshift samples, even when considering the error ranges, potentially hinting at the fact that the impact of environment in suppressing satellite star-formation increases as the Universe evolves. \textcolor{black}{Taking this further,} in the right panel of  Figure \ref{fig:SFSoffset} we then show the same, but split into our two different halo mass ranges. For the lower mass halos (gold lines) we find a similar $\sim$0.5\,dex log$_{10}$(SFR/M$_{\odot}$\,yr$^{-1}$) offset to all satellites,  while for the higher mass halos (blue lines) we see a stronger suppression of star formation with a  $\sim$0.85-1.1\,dex log$_{10}$(SFR/M$_{\odot}$\,yr$^{-1}$) offset at all redshifts. \textcolor{black}{Once again, details of the systematic SFR offsets between isolated centrals and satellites in these halo ranges are also calculated over the full 10$<$log$_{10}$(M$_{\mathrm{\star}}$/M$_{\odot}$)$<$11 range and given in Table \ref{tab:values}}. This also suggests that at a fixed stellar and halo mass the suppression of star formation in satellites is similar over the last $\sim$5\,Gyr, and that this suppression is stronger in higher mass halos. We also once again see the hint of the suppression becoming stronger as the universe evolves, but now also controlled for halo mass. \textcolor{black}{To highlight this, in Figure \ref{fig:timeOffset} we show the log$_{10}$(SFR/M$_{\odot}$\,yr$^{-1}$) offsets as a function of time for each of our halo mass ranges, given in Table \ref{tab:values}.  We then fit the time evolution using a simple linear regression fit. We find evolution of satellite SFR offsets, with all halo mass-selected samples increasing by $\sim$0.3\,dex towards the current epoch. We also see the clear separation of low- and high-mass halos in terms of log$_{10}$(SFR/M$_{\odot}$\,yr$^{-1}$). The offset increase with time is significant when considering the calculated errors, and suggests the impact of environmental quenching may be increasing with time (at the stellar masses probed here). }

\begin{figure}
\begin{center}
\includegraphics[scale=0.16]{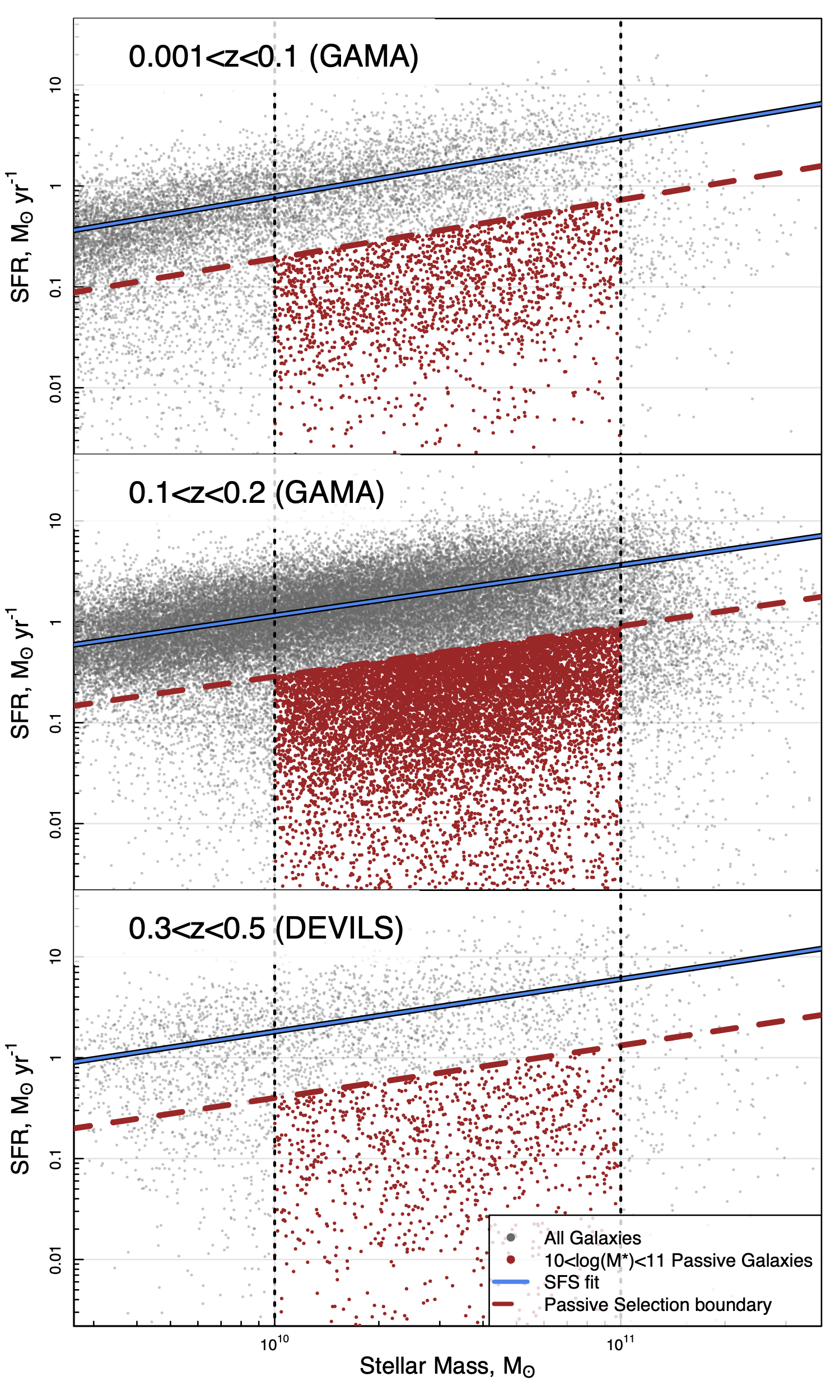}

\caption{The SFR-M$_{\star}$ plane at the three difference redshifts probed in this work showing the selection of passive galaxies. \textcolor{black}{At each redshift we identify the star-forming population and fit the locus of the SFS (blue line), we then define passive galaxies as falling $>$2$\sigma$ below this line (red  dashed line). This process is described in detail in Appendix \ref{sec:passive}}. In subsequent plots we only consider the passive fraction at $10<$log$_{10}$(M$_{\star}$/M$_{\odot}$)$<11$ bounded here by the vertical dashed lines. Sources selected as passive in this range are shown as red points.  }
\label{fig:PassiveSelection}
\end{center}
\end{figure}

\subsection{Evolution of the satellite passive fraction}

The previous results show that i) over-dense environments are impacting the star-formation in satellites, ii) that when controlled for stellar and halo mass that this effect has been weakly increasing over the last $\sim$5\,Gyr of Universal evolution and that iii) the suppression of star-formation is strongest in the most massive halos. This is largely consistent with our current understanding of galaxy evolution processes - that environment suppresses star formation in satellite galaxies by removing or heating star-forming gas, and that the strength of this suppression is largely dependant on halo mass (and more weakly with epoch). Whats more, this suppression of star-formation should lead to more and more galaxies transitioning from a star-forming to a passive quiescent state over time. While this occurs in all galaxies ($i.e.$ as traced by the decline in the cosmic star-formation history and downsizing), most importantly for this work, if it is the environment that is accelerating the transition through satellite quenching, we should expect to see the faction of passive galaxies in over-dense environments increasing more rapidly than in the field. 

To test this, we next consider the evolution of the passive fraction of satellite galaxies over the epochs probed here. To first select passive systems, we identify galaxies which sit significantly below the star-forming main sequence (SFS). Figure \ref{fig:PassiveSelection} shows this process \textcolor{black}{and it is described in detail in Appendix \ref{sec:passive}.  Briefly, we first identify the locus of the SFS by finding the running median SFR of the star-forming population, and performing a least squares regression to median points. These are given as the blue lines in Figure \ref{fig:PassiveSelection}. At each redshift, we then define a passive galaxy selection line that falls at 2$\sigma$ below this line - where $\sigma$ is the standard deviation of the star-forming population.}  All galaxies below this line are defined as passive, and are coloured red in Figure \ref{fig:PassiveSelection}. This process results in a selection line which is 0.61, 0.60 and 0.66 dex below the SFS in our increasing redshift bins respectively. 

\begin{figure*}
\begin{center}
\includegraphics[scale=0.8]{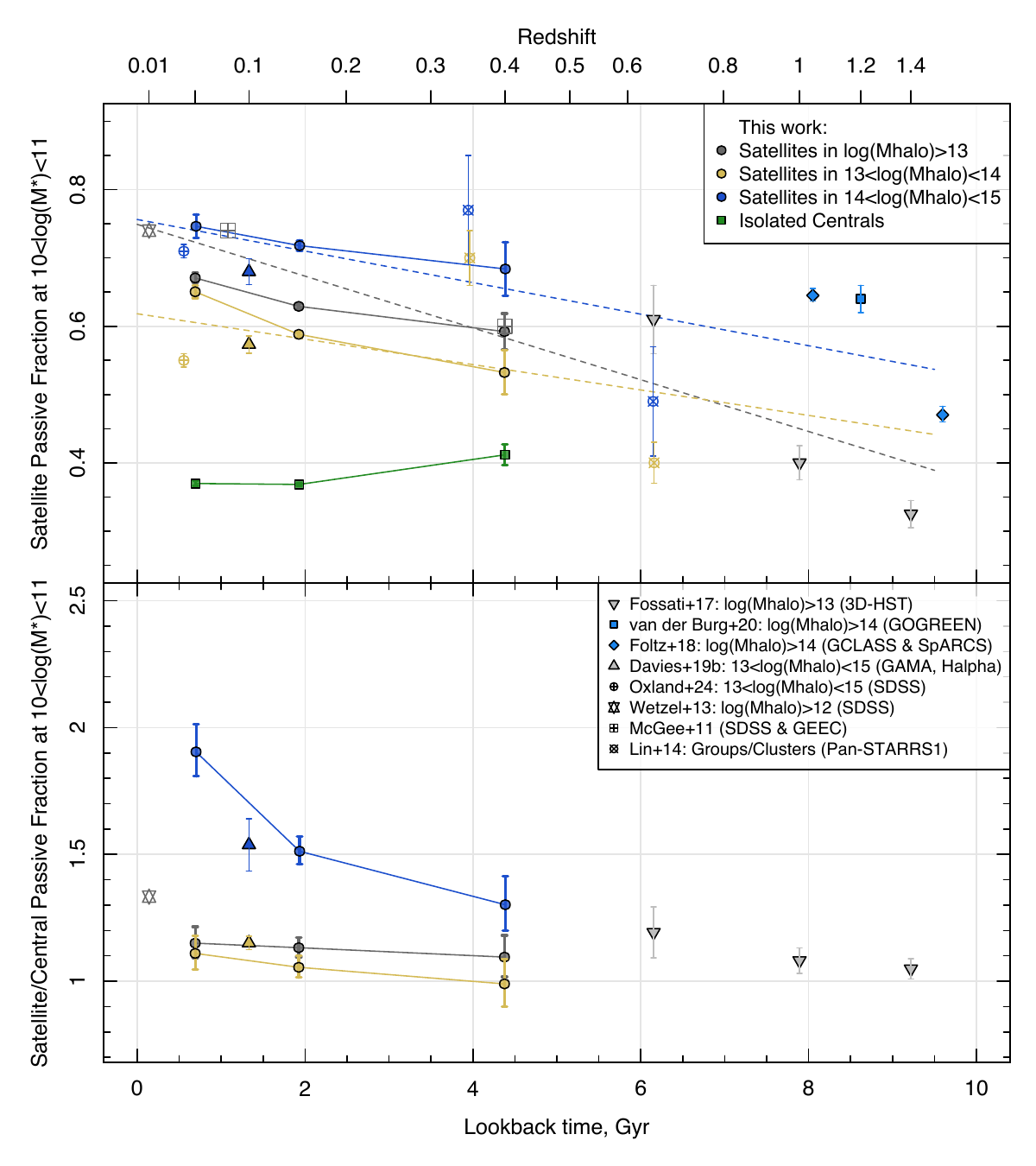}

\caption{Top: Evolution of the 10$<$log$_{10}$(M$_{\star}$/M$_{\odot}$)$<$11 passive fraction from $z$$\sim$1.5 to today showing a comparison between our current results and existing literature passive fractions. Points are coloured in halo mass ranges, such that points/lines of the same colours can be compared. Please see text for details of sample plotted. We also faintly show linear regression models to each halo mass range as dashed lines. Bottom: The same but showing (where samples allow) the halo satellite/central passive fraction.}
\label{fig:PassiveFractionComp}
\end{center}
\end{figure*}

Using this sample, we then calculate the fraction of passive galaxies at each epoch and display the time evolution of passive fractions for a number of sub-samples \textcolor{black}{in the top panel of} Figure \ref{fig:PassiveFractionComp} as filled points joined by lines. 1$\sigma$ confidences on each point are calculated from the binomial distribution estimated using a beta distribution following the procedure of \cite{Cameron11}. \textcolor{black}{We then combine these in quadrature with the error range of passive fractions for each sample derived from Monte-Carlo analysis outlined in Section \ref{sec:suppression}. As such, errors bars contain both statistical errors and fitting error in the measurement of SFR and stellar masses.  The evolution of passive fractions are shown for isolated centrals (green), satellites in all log$_{10}$(M$_{\mathrm{halo}}$/M$_{\odot}$)$>$13 halos (grey), 13$<$log$_{10}$(M$_{\mathrm{halo}}$/M$_{\odot}$)$<$14 halos (gold) and  14$<$log$_{10}$(M$_{\mathrm{halo}}$/M$_{\odot}$)$<$15  halos (blue). Data points from this work in Figure \ref{fig:PassiveFractionComp} are also provided in Table \ref{tab:fracs}.}

\begin{table*}
\caption{Passive fractions for 10$<$log$_{10}$(M$_{\mathrm{\star}}$/M$_{\odot}$)$<$11 galaxies and various samples calculated in this work. Error ranges are calculated using the range of offsets in the Monte-Carlo realisations described in the text, combined quadrature with the binomial distribution errors.}
\begin{center}
\begin{tabular}{ccc|c|cc|cc|cc }
& &  & & \multicolumn{2}{c}{log$_{10}$(M$_{\mathrm{halo}}$/M$_{\odot}$)$>$13} &  \multicolumn{2}{c}{13$<$log$_{10}$(M$_{\mathrm{halo}}$/M$_{\odot}$)$<$14} & \multicolumn{2}{c}{14$<$log$_{10}$(M$_{\mathrm{halo}}$/M$_{\odot}$)$<$15}  \\
Sample & $z$ & Lookback, Gyr & Isolated Centrals & Satellite & Central & Satellite & Central & Satellite & Central \\
\hline
 & & &  &  &  &  &  &  &  \\
GAMA & 0.001$<$$z$$<$0.1 & 0.70 & 0.37$\pm^{0.0042}_{0.0038}$ & 0.67$\pm^{0.008}_{0.0077}$ & 0.58$\pm^{0.01}_{0.0092}$ & 0.65$\pm^{0.0095}_{0.0091}$ & 0.59$\pm^{0.011}_{0.01}$ & 0.75$\pm^{0.018}_{0.017}$ & 0.39$\pm^{0.051}_{0.044}$ \\
 & & &  &  &  &  &  &  &  \\
GAMA & 0.1$<$$z$$<$0.2 & 1.93 & 0.37$\pm^{0.0021}_{0.002}$ & 0.63$\pm^{0.0043}_{0.0042}$ & 0.56$\pm^{0.0055}_{0.005}$ & 0.59$\pm^{0.005}_{0.0049}$ & 0.56$\pm^{0.0061}_{0.0056}$ & 0.72$\pm^{0.0078}_{0.0076}$ & 0.47$\pm^{0.019}_{0.016}$ \\
 & & &  &  &  &  &  &  &  \\
DEVILS & 0.3$<$$z$$<$0.5 & 4.38 & 0.41$\pm^{0.015}_{0.015}$ & 0.59$\pm^{0.026}_{0.026}$ & 0.54$\pm^{0.028}_{0.027}$ & 0.53$\pm^{0.032}_{0.032}$ & 0.54$\pm^{0.033}_{0.033}$ & 0.68$\pm^{0.04}_{0.039}$ & 0.53$\pm^{0.044}_{0.042}$ \\
\end{tabular}
\end{center}
\label{tab:fracs}
\end{table*}

 Here we first see that passive fractions for isolated centrals (green) are low and marginally decline over the epoch probed. In contrast, satellite galaxies in all halos (grey) increase moderately in passive fraction with time, \textcolor{black}{rising by $\sim10\%$ over the last 5\,Gyr. This is somewhat more pronounced} when we just consider the two halo mass ranges discussed in the previously, both 13$<$log$_{10}$(M$_{\mathrm{halo}}$/M$_{\odot}$)$<$14 halos (gold) and 14$<$log$_{10}$(M$_{\mathrm{halo}}$/M$_{\odot}$)$<$15  halos (blue) show an increase in passive fraction over the last $\sim5$\,Gyrs, \textcolor{black}{with the most extreme increase of $\sim12\%$ in passive fraction for lower mass halos}. This suggests that the impact of these halos is increasing the number of passive galaxies at these stellar masses - $i.e.$ environmental quenching.

We do note a word of caution that the process of selecting passive galaxies from the SFR-M$_{\star}$ plane is somewhat fraught with difficulty and subjective depending on the study at hand \citep[see][for discussion]{Davies25a,Davies25b} We have selected a method here that aims to be largely agnostic to the redshift evolution of the SFS in normalisation and scatter, and using similar metrics in each redshift bin (to allow comparisons of the passive fraction across epochs). However, we note that if different selection methods are used, different passive fractions are obtained. As such, comparing passive fractions across different studies can be problematic. However, we  briefly note that the passive fractions obtained for GAMA here for both satellites and isolated centrals, are consistent with passive fraction presented in other works (see next section).   

Finally, we also repeat our analysis using the much simpler method just selecting passive galaxies at just 1\,dex below the SFS at all epochs (which does not take into account the varying width of the SFS at a given epoch caused both variation in the intrinsic galaxy population and measurement errors), and find that our results hold true. The only difference in using this more simplistic method, is that the weak increase in satellite passive fraction between the 0.3$<$\textit{z}$<$0.5 bin and 0.1$<$\textit{z}$<$0.2 bin is removed, we still see an increasing satellite passive fraction to the 0.001$<$\textit{z}$<$0.1 bin. This highlights that care must be taken when defining passive fractions from the SFR-M$_{\star}$ plane, and any results should be interpreted with caution. 

That said, here we apply the same method of selecting passive galaxies for both satellites and isolated centrals and hence they should both be subject to the same methodology biases - somewhat negating the caveats noted above. As such, the potentially more robust result is the comparison of the difference between the evolution of the satellite populations and isolated centrals. Here we find that the isolated central ($i.e.$ galaxies with little or no environmental impact on their evolution) show a flat or declining passive fraction with time, while satellite galaxies show an increasing passive fraction.  This is true when using both methods for selecting passive galaxies noted above, suggesting that irrespective of methodology for identifying passive systems, over-dense environments are increasing in passive fraction more rapidly than isolated galaxies (at a fixed stellar mass) and therefore a galaxy's environment is driving its transition from a star-forming to a passive state.

\begin{figure*}
\begin{center}
\includegraphics[scale=0.72]{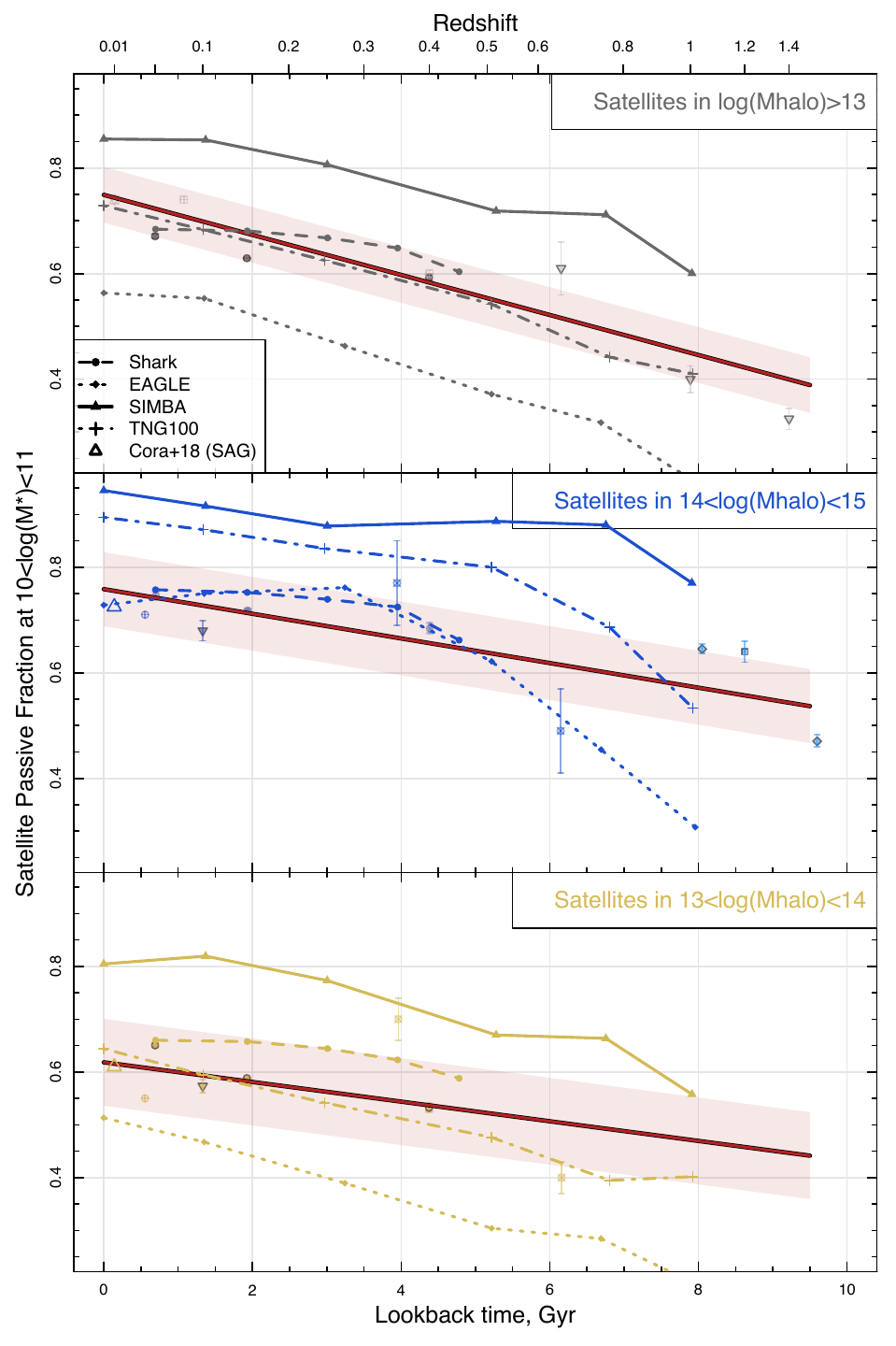}

\caption{Evolution of the 10$<$log$_{10}$(M$_{\star}$/M$_{\odot}$)$<$11 passive fraction from $z$$\sim$1.5 to today showing a comparison between our current results and numerical simulations. For ease of comparison, we split our halo mass ranges into three separate panels. Observational data is shown faintly in the background and to allow for more direct comparison, the red line in each panel shows a simple linear regression fit to the observational data. Different simulations are shown by different line types.  Points are coloured in halo mass ranges, such that points/lines of the same colours can be compared. Please see text for details of sample plotted.}
\label{fig:PassiveFractionComp_sim}
\end{center}
\end{figure*}

\subsubsection{Comparison to literature}

As noted above, comparing passive fractions across different studies/samples is fraught with difficulty as each work uses different selections and methodologies which can bias any direct comparison of passive fractions. However, in this section we aim to compare our satellite passive fractions with a number of other studies and, where possible, identify literature results which cover comparable stellar and halo mass ranges. These comparisons are summarised in Figure \ref{fig:PassiveFractionComp}. 

For the local Universe we first compare to our own satellite passive fractions from GAMA at $z$$<$0.2 and matched to the halo and stellar mass ranges probed here \citep{Davies19b}. \textcolor{black}{We explicitly re-calculate the \cite{Davies19b} points using the exact stellar and halo mass ranges used in the current work}. These are shown as downwards triangles, coloured to match the two halo mass range used here. These passive fractions use the same group catalogues as here, but a selection of star-forming/passive galaxies using H$_{\mathrm{alpha}}$-based SFRs (instead of SED-based SFRs).  \textcolor{black}{We first find that the difference in passive fraction between the low- and high-mass halos is consistent between \citep{Davies19b} and our current work. However, we do find a small, but significant, normalisation offset between the two samples. This is likely due to differences in the selection of passive galaxies, which is discussed further below.}  Next we take the $z$$<$0.1 satellite passive fractions derived in \cite{Oxland24} for the Sloan Digital Sky Survey (SDSS). \cite{Oxland24} calculate passive fraction as a function of time since satellite infall for a number of stellar mass ranges, and over the same halo mass ranges we use here. We use their passive fractions for 10$<$log$_{10}$(M$_{*}$/M$_{\odot}$)$<$11 galaxies  as $0.71\pm0.01$ and $0.55\pm0.01$ for high and low mass haloes respectively (via private communication). These are shown as the crossed circles in of Figure \ref{fig:PassiveFractionComp}, which are also lower than our current measurements at $z$$\sim$0. Next we show the earlier work of \cite{Wetzel13} also using SDSS, where we take the log$_{10}$(M$_{*}$/M$_{\odot}$)$=$10.5 point from their figure 4. This includes satellite in all log$_{10}$(M$_{\mathrm{halo}}$/M$_{\odot}$)$>$12 halos at $z$$\sim$0 identified in that work. This is shown as the grey star in Figure \ref{fig:PassiveFractionComp} . The grey crossed squares show the passive fractions from \cite{McGee11}, who calculate passive fractions for SDSS at $z$$\sim$0.08 and the Group Environment Evolution Collaboration (GEEC) at $z$$\sim$0.4. We take the log$_{10}$(M$_{*}$/M$_{\odot}$)$=$10.5 points for group galaxies from their figure 8. For SDSS samples both \cite{Wetzel13} and \cite{McGee11} both have higher passive fractions than our current analysis. The GEEC point from \cite{McGee11} is almost identical to our passive fraction at earlier times. Lastly, we also show the Pan-STARRS 1 passive fractions presented in \cite{Lin14}, taken directly from their table 2 for log$_{10}$(M$_{*}$/M$_{\odot}$)$=$10.45 satellites in groups (gold circle with cross), and in clusters (blue circle with cross).  We note that their passive fraction is significantly higher than our results at $z$$\sim$0.35, but declines rapidly to earlier lookback times.  However, we also highlight that the \cite{Lin14} work uses purely photometric samples, and is therefore likely contaminated and/or biased in group/cluster identification and satellite assignment - complicating the measurement of passive fractions.           

Next, extending to higher redshifts we first include passive fractions from the 3D-HST analysis of \cite{Fossati17}. They match to mock galaxy catalogues to obtain probabilistic halo masses and select star-forming/passive galaxies based off broad-band colours. As such, we note that this work has vastly different methodology to the one we use here.  From figure 17 of \cite{Fossati17} we extract the log$_{10}$(M$_{\mathrm{halo}}$/M$_{\odot}$)$>$13 points at log$_{10}$(M$_{*}$/M$_{\odot}$)$=$10.5 from each of the right panels at the three epochs probed. These are displayed as the light grey downwards triangles in Figure \ref{fig:PassiveFractionComp}. Despite the vastly different methodologies, these point show similar passive fractions to our data, and a declining passive fraction with lookback time. However, the lowest redshift point from \cite{Fossati17} appears more consistent with our most massive halos instead of our full halo sample, which should cover the same halo mass range. Next we include a number of passive fractions from satellites in massive, cluster scale halos at high redshift. \cite{vanderBurg20}  calculate passive and star-forming galaxy stellar mass functions for cluster galaxies at $z$$\sim$1.2 from the GOGREEN survey \citep{Balogh17}. Here we extract the log$_{10}$(M$_{*}$/M$_{\odot}$)$=$10.45 passive fraction from their table 3. This is displayed as the   blue square in Figure \ref{fig:PassiveFractionComp}. This is largely consistent with our highest redshift point for the most massive halos (slightly higher). Lastly, \cite{Foltz18} explore the satellite passive fraction in massive SpARCS \citep{Muzzin09} clusters at $z$$>$1. They provide their red, green and blue satellite numbers at two different redshifts. Here we calculate red/passive fraction and errors using the same methodology as in our GAMA/DEVILS sample.  These are shown as the blue diamonds in Figure \ref{fig:PassiveFractionComp} and have a similar massive fraction to \cite{vanderBurg20}  and our highest mass halos.

Here, overall we see a qualitative general of passive fractions increasing with Universal time in a largely similar way in each of our halo mass ranges and consistency between our current results and the existing literature. In each of the halo mass ranges, we see a roughly linear increase in satellite passive fraction over time, with passive fractions increasing by a $\sim$0.01-0.03 per Gyr. While the evolutionary trends are similar the normalisation of absolute passive fraction is higher for the most massive halos.  

\textcolor{black}{As noted above, comparing literature results in this way is fraught with difficulty due to the different methodologies used by different teams when defining both environment and passive galaxies. A potential solution to remove some of the biases in selection of passive galaxies is to compare to the evolution of satellite-to-central passive fractions - which should be subject to the same methodology choices. In the bottom panel of Figures \ref{fig:PassiveFractionComp}, we show the time evolution of the satellite-to-central passive fraction for our sample. We find that lower mass halos only marginally increase in satellite-to-central passive fraction over the last $\sim$5\,Gyrs, while in the most massive halos satellite-to-central passive fraction increase significantly rising by $\sim$85\%. This is consistent with the picture of the most massive halos having the most significant impact on the star-formation in satellite galaxies, with satellites evolving more significantly than centrals in these systems. }   

\textcolor{black}{Only \cite{Wetzel13}, \cite{Fossati17} and \cite{Davies19b} provide the information with which to calculate satellite-to-central passive fractions, and as such only these works are displayed in the bottom panel of Figure \ref{fig:PassiveFractionComp}. Firstly, considering the satellite-to-central passive fractions for all  log$_{10}$(M$_{\mathrm{halo}}$/M$_{\odot}$)$>$13 halos (grey) we see qualitative agreement with both \cite{Wetzel13} and \cite{Fossati17}, with consistent satellite-to-central passive fractions around 1.0-1.3 and a weak evolution with time, increasing by a factor of $\sim30\%$ over the last $\sim$9\,Gyrs. Only \cite{Davies19b} allows for the satellite-to-central passive fraction to be split by halo mass, and we see broad agreement with the satellite-to-central passive fractions found in our current work, with a similar offset in satellite-to-central passive fraction between high and low mass haloes  -  0.41 for \cite{Davies19b} and 0.58 for this work (interpolated at the same lookback time). This potentially suggests that the differences in satellite passive fraction described in the top panel of Figure \ref{fig:PassiveFractionComp}, are in fact due to different passive galaxy selection methods.}

\subsubsection{Comparison to simulations}

In Figure \ref{fig:PassiveFractionComp_sim} we compare all of these observational results to a number of different simulations. For clarity, here we split out each of our halo mass ranges into separate panels. In this figure the observational points as each halo mass range are shown faintly in the background. To allow for more direct comparison, the red line in each panel shows a simple linear least-squares regression fit to the observational data \textcolor{black}{(these are also shown as the dashed lines in the top panel of Figure \ref{fig:PassiveFractionComp}). The red polygon show the 1$\sigma$ orthogonal offset from this line for all observational datasets.} Different simulations are shown with different point and line types.   For consistency between simulations, we use an identical method for selecting passive galaxies in each sample following \cite{Wright22}, who select passive galaxies as having log$_{10}$(sSFR/yr$^{-1}$)$<$$-11+0.5z$ (irrespective of the SFS). We then use the simulation snapshots at a range of epochs, and select halo and stellar mass ranges comparable to our observations. This is first plotted for the  \textsc{Shark} semi-analytic model \citep{Lagos18, Lagos24} as the dashed lines with dot points. We then include a number of hydrodynamical simulations determined in an identical manner: dotted lines with diamond points for EAGLE \citep{Crain15, Schaye15}, solid lines with triangle points for SIMBA \citep{Dave19} and dot-dashed lines with cross points for TNG100 \citep{Pillepich18, Nelson19}. For further detailed discussion of satellite quenching in TGN100 and EAGLE see \cite{Donnari21a, Donnari21b} and \cite{Wright22} respectively. Finally, in the middle and bottom panel we also show the $z\sim0$ passive fractions from the semi-analytic model of galaxy formation, SAG, applied on the
MULTIDARK simulation presented in \cite{Cora18}. From their figure 11 we extract their quenched fraction for log$_{10}$(M$_{*}$/M$_{\odot}$)$=$10.5 galaxies in both 14$<$log$_{10}$(M$_{\mathrm{halo}}$/M$_{\odot}$)$<$15 and 13$<$log$_{10}$(M$_{\mathrm{halo}}$/M$_{\odot}$)$<$14 halos.

\begin{figure*}
\begin{center}
\includegraphics[scale=0.75]{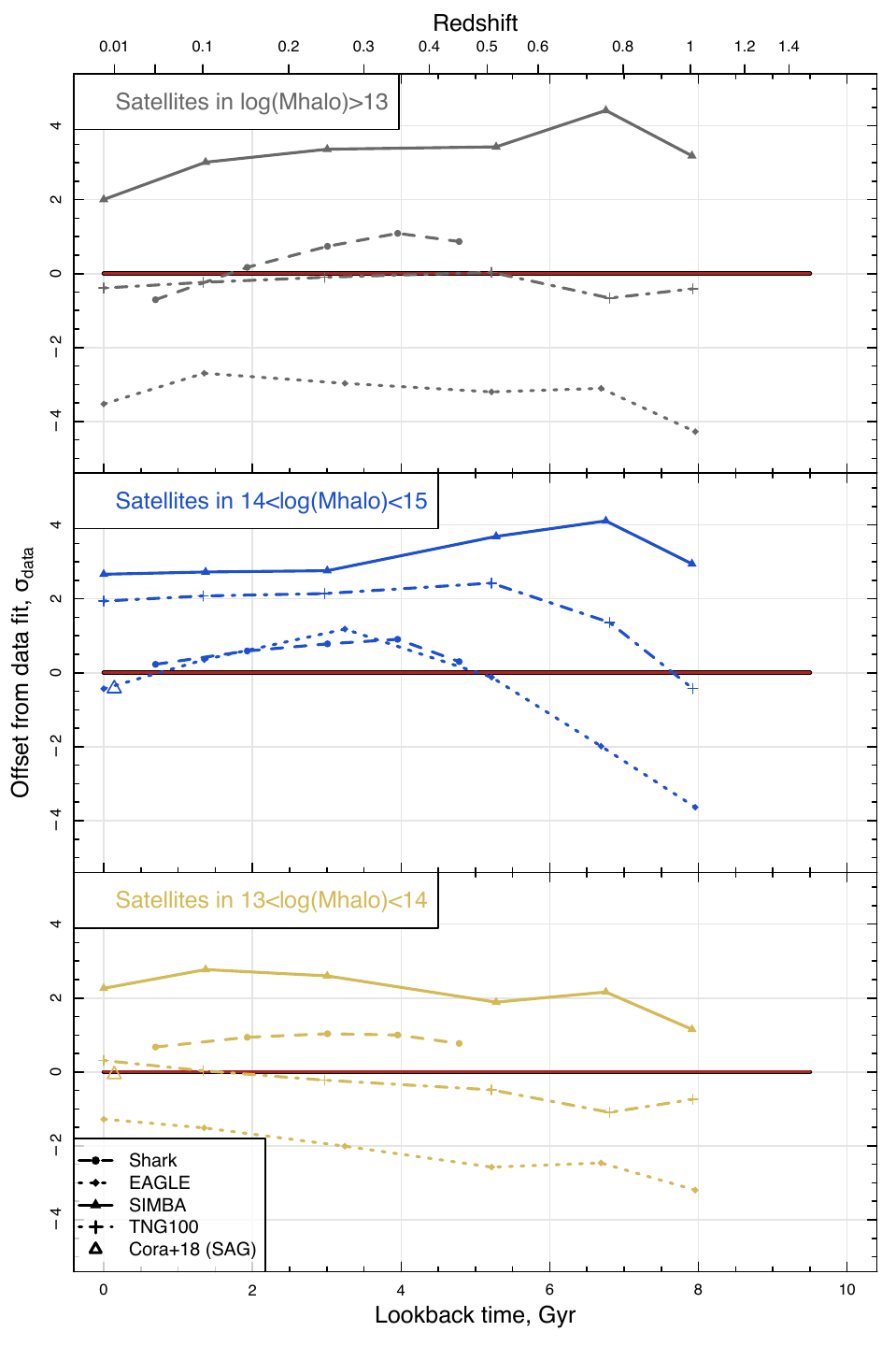}

\caption{Similar to Figure \ref{fig:PassiveFractionComp_sim} but showing the offset of each of the simulations from the observational trend, in units of 1$\sigma$ spread of the observational data. Simulations that sit along the 0 line are consistent with the observational trends, while offset from this line shows $\sigma$ tension with the observational trends}
\label{fig:PassiveFractionComp_sim_offset}
\end{center}
\end{figure*}

\textcolor{black}{To directly compare these simulation trends with the observational datasets, in Figure \ref{fig:PassiveFractionComp_sim_offset} we show panels similar to Figure \ref{fig:PassiveFractionComp_sim}, but now as the offset from the fit to the observational data, scaled by the 1$\sigma$ spread of the observational data. As such, in this figure, a simulation at zero matches the observational trends exactly. Deviations from this show $\sigma$ tension with the observational trends derived in this work.       }

\textcolor{black}{Within Figures \ref{fig:PassiveFractionComp_sim} and \ref{fig:PassiveFractionComp_sim_offset}, we find that, interestingly, all of the simulations show very similar evolutionary trends to the observational data (slopes) but differ quite significantly in normalisation.  For example, SIMBA has much higher passive fractions than the observations falling at 2-3 $\sigma$ above the observational trends, but has a very similar evolutionary slope. This is systematically true for all halo masses. Likewise, EAGLE also shows a similar evolutionary trend, but has much lower normalisation (2-3 $\sigma$ below the observational trends) when considering all but the most massive halo. However, TNG 100 appears to match the observational data very well, for lower mass halos, having very little tension with the observational trends for both the full sample and lower mass halos. SAG at $z\sim0$ is very well-matched to the observation data at both halo mass ranges. Finally, \textsc{Shark} matches the observational trends well at all halo masses, showing at most 1\,$\sigma$ tension for the low halo mass sample.}

 \textcolor{black}{These differences in normalisation for the hydrodynamical simulations are likely driven by the specifics of the feedback physics models used \citep[$i.e.$ similar SFR offsets between simulations exist for all galaxies][]{Haidar22}. For example, SIMBA uses a strong AGN feedback model which leads to overall higher passive fractions in all galaxies. Likewise, the SAG and \textsc{Shark} semi analytic models may differ due to how each simulation is calibrated - with SAG being tuned based on the $z$$\sim$0 SFR function, while \textsc{Shark} is not. Therefore, SAG is likely to better match the $z$$\sim$0 star-forming properties of galaxies by design.} However, we note that a detailed discussion of the evolution of passive fractions in simulations is beyond the scope of this work. Here we simply aim to show that our current observational results are consistent with contemporary simulations, within the spread induced by their choice of underlying astrophysics and methodology, and note that based off the observations presented here,  \textsc{Shark} and TNG 100 represent the best match to the observational trends.

\section{Summary and Conclusions}

In this work we have presented the evolution of the star-forming properties of satellite galaxies over the last $\sim$5\,Gyr ($z$$<$0.5) using the DEVILS and GAMA samples. Both DEVILS and GAMA use the same methodology for measuring galaxy properties and defining environmental properties, thus reducing any biases induced in using different methodologies across different samples. This allows us to define a robust stellar mass and halo mass complete sample across all of the epochs studied. Using these samples we show that, at a fixed stellar mass, satellite galaxies have significantly lower SFRs than isolated central galaxies at all epochs \textcolor{black}{(by $\sim$0.5\,dex in log$_{10}$(SFR/M$_{\odot}$\,yr$^{-1}$))}, and that this suppression of star formation increases over the last $\sim$5\,Gyr. We then show that the strength of this suppression also scales with halos mass, with the most massive dark matter halos showing the most highly suppressed satellites\textcolor{black}{, of up to $\sim$1\,dex in log$_{10}$(SFR/M$_{\odot}$\,yr$^{-1}$) for the most massive halos}. Once again, we see that this effect is largely consistent across all epoch probed. This suggests that satellite galaxies are having their star-formation suppressed (quenching) at all epochs and that the strength of environmental quenching processes is largely the same at all epochs, when controlled for both stellar mass and halo mass. Next we identify passive galaxies, based on position relative to the star-forming sequence, and show that, as expected, passive fractions are higher in group/cluster environments than in isolated centrals at the same stellar mass, with passive satellite fractions up to twice as large as their isolated counterparts. We find that satellite passive fractions increase with time and at a far faster rate than in isolated centrals, and that this effect is strongest in the most massive halos. This suggests that most over-dense environments are driving the transition of galaxies from an actively star-forming to a passive/quiescent state - consistent with our current understanding of galaxy evolution processes. 

We note that comparing passive fractions across different studies and their evolution is problematic due to measurement and methodology differences. However, we do compare our current results with existing literature measurements of satellite passive fractions over the epochs and halo masses probed here, and find that they are largely consistent, showing similar absolute passive fractions and a similar overall evolutionary trend. Finally, we also compare to modern galaxy evolution simulations, and find largely similar \textcolor{black}{evolutionary slopes of satellite passive fractions but sometimes wildly varying normalisations - likely due to the different feedback mechanisms employed.}  

\section*{Acknowledgements}

LJMD and MFFF acknowledge support from the Australian Research Council's Future Fellowship scheme (FT200100055). LJMD acknowledges support from the Australian Research Council's Discovery Project scheme (DP250104611). RJW acknowledges the generous support of the Forrest Research Foundation. MB is funded by McMaster University through the William and Caroline Herschel Fellowship. M.S. acknowledges support by the State Research Agency of the Spanish Ministry of Science and Innovation under the grants 'Galaxy Evolution with Artificial Intelligence' (PGC2018-100852-A-I00) and 'BASALT' (PID2021-126838NB-I00) and the Polish National Agency for Academic Exchange (Bekker grant BPN/BEK/2021/1/00298/DEC/1). This work was partially supported by the European Union's Horizon 2020 Research and Innovation program under the Maria Sklodowska-Curie grant agreement (No. 754510). DEVILS is an Australian project based around a spectroscopic campaign using the Anglo-Australian Telescope. DEVILS is part funded via Discovery Programs by the Australian Research Council and the participating institutions. The DEVILS website is \url{devilsurvey.org}. The DEVILS data are hosted and provided by AAO Data Central (\url{datacentral.org.au}).


\vspace{-5mm}

\section{Data Availability}

Data products used in this paper are taken from the internal DEVILS
team data release and presented in \cite{Davies21} and \cite{Thorne21}. These catalogues will be made public as part DEVILS
first data release described in Davies et al. (in preparation). GAMA data is taken from the GAMA DR4 public archive: \url{https://www.gama-survey.org/dr4/schema/}

\appendix

\section{Selection of passive galaxies}
\label{sec:passive}

\begin{figure*}
\begin{center}
\includegraphics[scale=0.2]{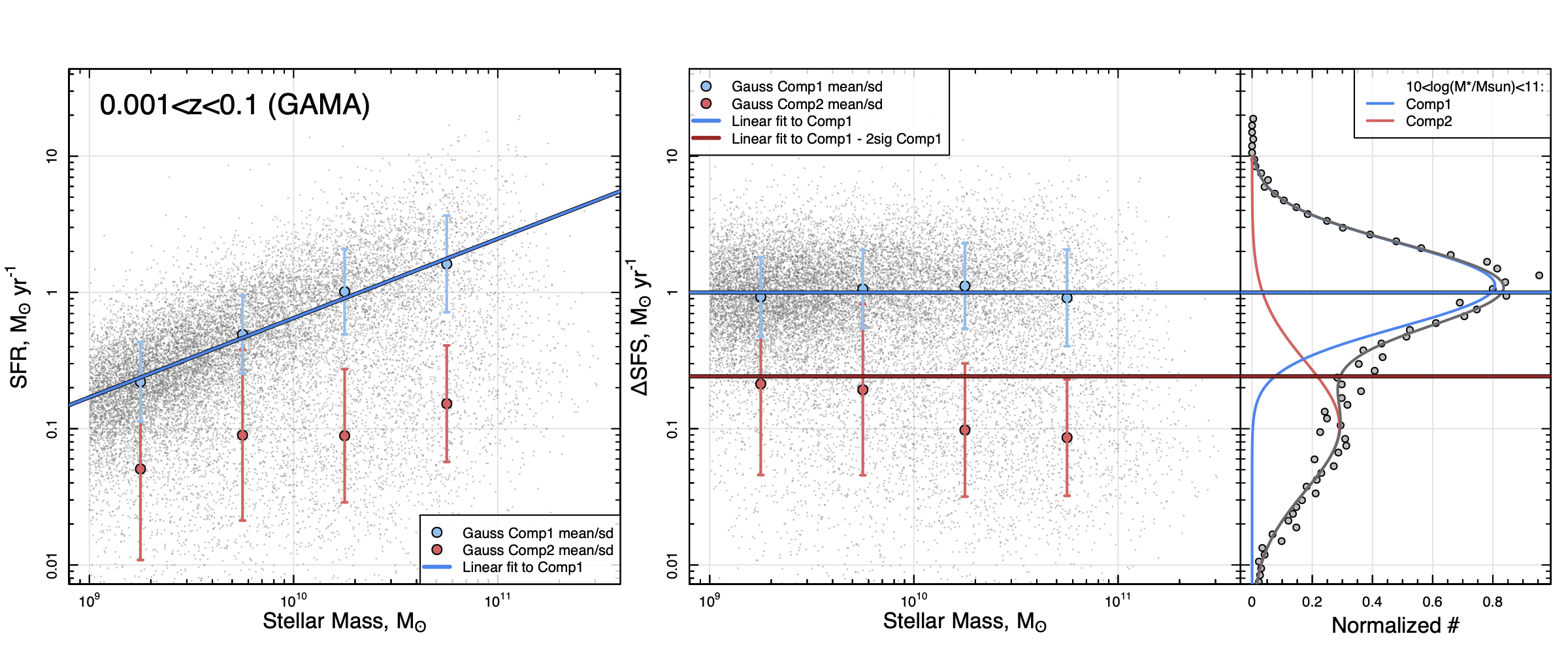}
\includegraphics[scale=0.2]{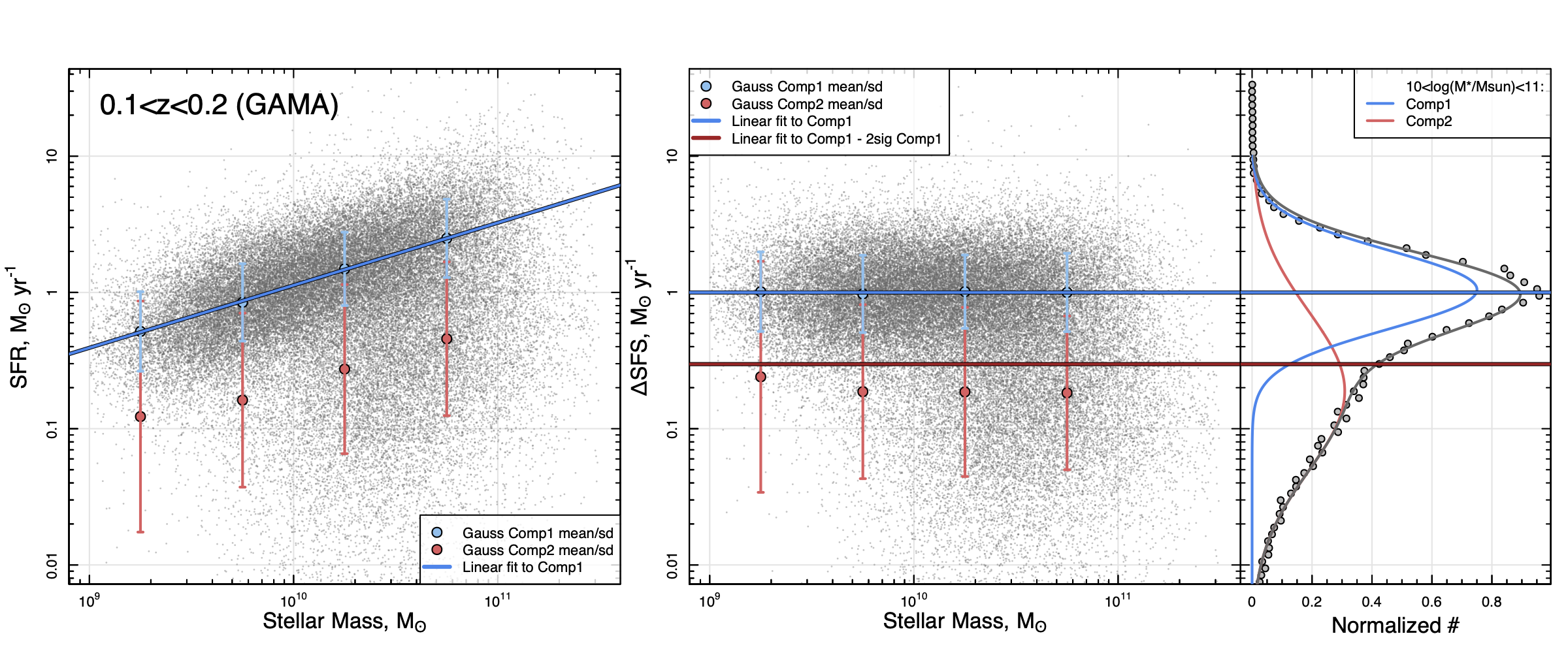}
\includegraphics[scale=0.2]{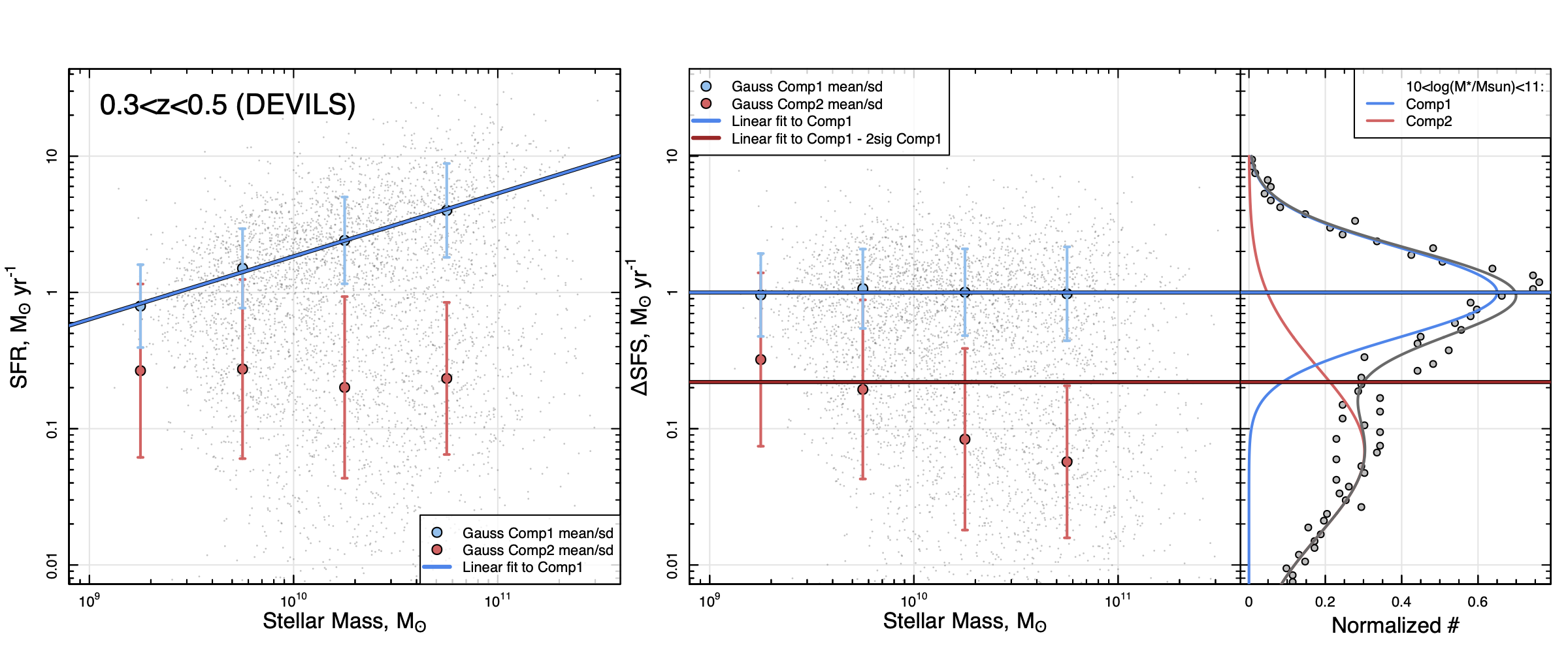}
\caption{Methodology for selecting passive galaxies at each epoch. Left: the SFR-M$_{\star}$ plane. In $\Delta$log$_{10}$(M$_{\star}$/M$_{\odot}$)=0.5 bins between 9$<$log$_{10}$(M$_{\star}$/M$_{\odot}$)$<$11 we apply a Gaussian mixture model to log$_{10}$(SFRs) to define the star-forming and passive populations. Means and standard deviations of the Gaussian components are shown as the coloured points and error bars respectively. The blue line shows the fit to the star-forming population means ($i.e.$ the SFS). Middle: the same but with the SFS slope and normalisation removed giving offset from the SFS, $\Delta$SFS. Right: The histogram of $\Delta$SFS for 10$<$log$_{10}$(M$_{\star}$/M$_{\odot}$)$<$11 galaxies, with the Gaussian mixture model overlaid. We define our passive galaxy selection line at 2$\sigma$ of the star-forming population below the SFS. These are then used in Figure \ref{fig:PassiveSelection}.      }
\label{fig:offsets}
\end{center}
\end{figure*}

\textcolor{black}{In this section we detail the methodology used to define the passive galaxies, summarised in Figure \ref{fig:PassiveSelection}. Details of this process are outlined in Figure \ref{fig:offsets}. First, at each redshift used in this work, we take the SFR-M$_{\star}$ plane, as shown in the left panel of Figure \ref{fig:offsets}. Then in bins of $\Delta$log$_{10}$(M$_{\star}$/M$_{\odot}$)=0.5 between 9$<$log$_{10}$(M$_{\star}$/M$_{\odot}$)$<$11, we apply a 2-component Gaussian mixture model to the distribution of log$_{10}$(SFRs) using the \texttt{R} \texttt{mixtools:normalmixEM} function. We set starting parameters of component mixing ratio, lambda=0.25, means=[0.75, 0] and standard deviations=[0.1,0.1]. Resultant fitted means and standard deviations for the components are shown in the left panel of Figure \ref{fig:offsets}. We then define the higher SFR component (blue points) as the star-forming population, and apply a linear regression model to fit the component means (blue line)- hereafter the SFS. We then remove the slope and normalisation of the SFS to derive the offset of each galaxy from the SFS, $\Delta$SFS (shown in the middle panel of Figure \ref{fig:offsets}). We then take all galaxies at 10$<$log$_{10}$(M$_{\star}$/M$_{\odot}$)$<$11 (the sample used in the main body of the paper), and show the distribution of $\Delta$SFS in the right panel of Figure \ref{fig:offsets}. We then refit this sample with the 2-component Gaussian mixture model to parametrise the star-forming and passive population for the sample used in this work, shown as the coloured lines in the right panel of Figure \ref{fig:offsets}. This is largely undertaken to define the standard deviation, $\sigma$, of the star-forming population. We then define the passive galaxy selection line at 2$\sigma$ below the SFS at each epoch - shown as the red line in the middle and right panels. The blue and red lines in this figure are the lines shown in Figure \ref{fig:PassiveSelection}. }                 

\bsp	
\label{lastpage}
\end{document}